\documentclass[a4paper,fleqn,usenatbib,useAMS]{mnras}

\usepackage{graphicx}	
\usepackage{amsmath}	
\usepackage{amssymb}	
\usepackage{multicol}        
\usepackage{multirow}
\usepackage{bm}		
\usepackage{pdflscape}	

\usepackage[T1]{fontenc}
\usepackage{ae,aecompl}

\title[Europium production by neutron star mergers]{Semi-analytic modelling of the europium production by neutron star mergers in the halo of the Milky Way}

\author[P. van Oirschot et al.]{Pim van Oirschot$^{1}$\thanks{e-mail: P.vanOirschot@astro.ru.nl},
Gijs Nelemans$^{1,2}$, Onno Pols$^1$ \& Else Starkenburg$^3$ 
\\
$^1$Department of Astrophysics/IMAPP, Radboud University Nijmegen, P.O. Box 9010, 6500 GL Nijmegen, The Netherlands \\
$^2$Institute for Astronomy, KU Leuven, Celestijnenlaan 200D, 3001 Leuven, Belgium\\
$^3$Leibniz-Institut f\"ur Astrophysik Potsdam, AIP, An der Sternwarte 16, D-14482 Potsdam, Germany}

\pubyear{2017}

\begin{document}
\label{firstpage}
\pagerange{\pageref{firstpage}--\pageref{lastpage}}
\maketitle

\begin{abstract}
Neutron star mergers (NSM) are likely to be the main production sites for the rapid (r-) neutron capture process elements.
We study the r-process enrichment of the stellar halo of the Milky Way through NSM, by tracing the typical r-process 
element Eu in the Munich-Groningen semi-analytic galaxy formation model, applied to three high resolution 
Aquarius dark matter simulations. In particular, we investigate the effect of the kick velocities that neutron 
star binaries receive upon their formation, in the building block galaxies (BBs) that partly formed the 
stellar halo by merging with our Galaxy. When this kick is large enough to overcome the escape velocity of
the BB, the NSM takes place outside the BB with the consequence that there is no r-process enrichment.
We find that a standard distribution of NS kick velocities decreases [Eu/Mg] abundances of halo stars by 
$\sim 0.5$~dex compared to models where NS do not receive a kick. 
With low NS kick velocities, our simulations match observed [Eu/Mg] abundances of halo stars reasonably well, 
for stars with metallicities [Mg/H]$\geq -1.5$. 
Only in Aquarius halo B-2 also the lower metallicity stars have [Eu/Mg] values similar to observations.
We conclude that our assumption of instantaneous mixing is most likely inaccurate for modelling the r-process enrichment 
of the Galactic halo, or an additional production site for r-process elements is necessary
to explain the presence of low-metallicity halo stars with high Eu abundances. 
\end{abstract}

\begin{keywords}
Galaxy:abundances, Galaxy:evolution, Galaxy:halo, stars:neutron
\end{keywords}
%
%

\section{Introduction}
The rapid (r-) neutron capture process which leads to the formation of roughly half of the elements heavier than iron
has been known since the classical paper of \citet{Burbidge:1957}. Yet, there is an ongoing debate about its
astrophysical production sites. The two main candidates are neutron star mergers (NSM) and core-collapse supernovae (SNe), 
where for the latter scenario a distinction is made between the prompt explosions of massive 
stars with masses in the range $8 - 10 \ M_\odot$ \citep[eg.][]{Wheeler:1998} and the delayed explosions of very massive stars 
\citep[$\geq 20 \ M_\odot$, eg.][]{Woosley:1994}. 
Those with a mass between 10 and 20 $M_\odot$ are not likely candidates because chemical evolution
models predict little to no scatter in [Eu/Fe] abundance ratios for the stars born out of their ashes, which clearly contradicts observations \citep{Ishimaru:1999aa}.
Whereas neutrino-driven winds associated with delayed explosions 
seem too proton rich to be the production sites of the high-mass r-process elements \citep{Arcones:2007,Arcones:2013}, 
prompt explosions might not actually exist in reality \citep{Argast:2004}.
Other suggested r-process production sites are long-duration gamma-ray bursts \citep[see e.g.][]{Metzger:2008aa}, 
magnetic proto-neutron star winds \citep{Suzuki:2005aa} and supernovae fallback \citep{Fryer:2006aa} amongst others. 

This paper focusses on the NSM scenario for r-process enrichment, which is arguably the most likely r-process 
production site, because of the following reasons:
\begin{itemize}
\item NSM nucleosynthesis models robustly produce both heavy and light r-process nuclei \citep{Goriely:2011,Wanajo:2014}.
\item The ${}^{244}$Pu interstellar medium density as a function of time is incompatible with continuous r-process production, 
as would be the case in the core-collapse SNe production scenario, but instead points towards a 
low rate / high yield production site, and perfectly matches the NSM scenario \citep{Wallner:2015,Hotokezaka:2015}.
\item The recent direct detection of gravitational waves from a NSM and
associated signal emitted by radioactive (r-process) material \citep{Abbott:2017aa,Abbott:2017ab}
 favor the NSM scenario for r-process enrichment.
\item Ultra faint dwarf galaxies and the Milky Way are likely to have the same production mechanism for r-process elements, 
triggered by rare events \citep{Beniamini:2016ab}.
\end{itemize}
However, a combination of the two different production sites could also explain the
origin of r-process elements in our Galaxy \citep[eg.][]{Ji:2016a}.

\begin{figure}
 \includegraphics[width=\columnwidth]{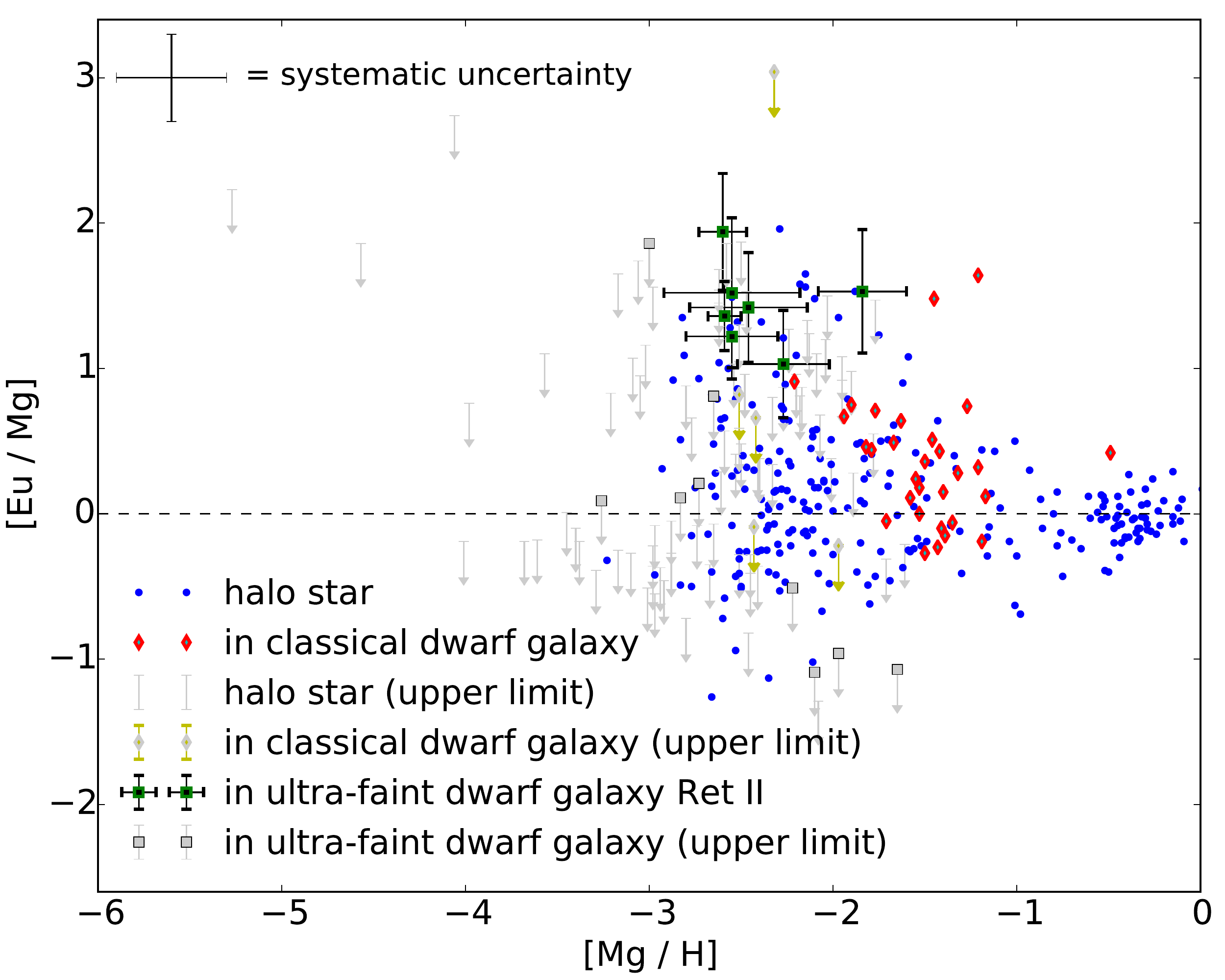}    
 \caption{[Eu/Mg] abundances for halo stars selected from the VizieR database \citep{Frebel:2010}.
 The typical errorbar of $\sim 0.3$ dex in abundance analyses is indicated in the upper left corner.
 Halo stars are shown as blue dots, red diamonds represent stars originating in classical dwarf galaxies
 and symbols with downward pointing grey arrows indicate upper limits. 
 The r-process enhanced stars in the ultra-faint dwarf galaxy Ret II \citep{Ji:2016,Ji:2016a} 
 are not included in the VizieR database, but are manually added to this figure 
 as green squares with errorbars.
 }
 \label{fig:1}
\end{figure}

We study the r-process enrichment by using europium (Eu) as its tracer, because this stable
r-process element is one of the r-process elements that can be observationally detected without 
exceptionally high data quality \citep[eg.][]{Frebel:2010}.
In Figure~\ref{fig:1} we show the scatter in [Eu/Mg] as a function of [Mg/H], for stars in the Galactic halo
and in nearby dwarf galaxies, made after querying the VizieR database \citep{Frebel:2010}. 
Mg is chosen as a metallicity tracer to compare the Eu-abundance with, instead of the more commonly used Fe, 
to obtain an abundance distribution that is directly comparable to our model predictions (as we will explain in the next sections). 
At low metallicities (eg. [Mg/H]$<-2.5$) some halo stars have high Eu abundances, a feature that
should be explained by any chemical evolution model of the r-process.

For many years it has been assumed that NSM alone are unable to reproduce the [Eu/Mg] ratios for stars with 
[Mg/H]$<-2.5$ \citep{Argast:2004,Wehmeyer:2015}, i.e. that Eu enrichment through NSM cannot take place at these 
low metallicities, mainly because NSM would be too rare and their average coalescence times too large. 
Also \citet{Matteucci:2014} and \citet{Cescutti:2015} showed that neutron star binaries can only be the dominant 
production sites of r-process elements if they merge after 1 Myr or in less than 10~Myr, respectively.
However, population synthesis studies of neutron star binaries suggest a wide distribution of delay times proportional to $1/t$,
with the first NSM occurring after 10~Myr or 100 Myr and $t$ the time since the formation of the neutron star binary \citep{Belczynski:2006}.
However, a small fraction of double neutron stars can merge on even smaller timescales of 1-10 Myr
\citep[e.g.][]{Beniamini:2016aa}.

Recently, \citet{Tsujimoto:2014}, \citet{Ishimaru:2015} and \citet{Komiya:2016aa} have shown that Eu-enhanced halo stars 
are expected in a Galactic halo that is formed from merging subhaloes, where the Eu originates from NSM with a coalescence time 
of $\sim$100~Myr. There is only a small probability that the subhaloes with small galaxies ($\sim 10^5 M_\odot$)
host a NSM at early times, but if this occurs the next generation of stars formed in such a galaxy would be
extremely enhanced in Eu (eg. [Eu/Mg]$\gtrsim +2$), because of the large Eu yields per NSM.
Therefore, this can be considered an other argument in favor of the NSM production scenario
for r-process elements, as it is now generally accepted that the Galactic halo
has in part grown from the mergers of smaller systems, building blocks (BB) as it were, 
a scenario that is based on the hierarchical growth of structures as explored in e.g. \citet{White:1978aa}. 
Observational evidence supporting growth via accretion and mergers has been reported in e.g.
\citet{Ibata:1994}, \citet{Helmi:1999a} and \citet{Belokurov:2006} amongst others. 

However, with the exception of the work of \citet{Beniamini:2016aa, Beniamini:2018aa}, and \citet{Hotokezaka:2018ab}, 
little attention has been given in the literature to the fact that neutron star binaries receive a kick velocity at the 
formation time of the two neutron stars, which might expel the system out of its host galaxy at early times, when the escape velocity 
of the host galaxy is still small. Therefore, we investigate in this paper if it is still possible to enrich 
the early Milky Way galaxy in r-process elements through the NSM scenario if neutron star kicks are taken into 
account. We use a semi-analytic galaxy formation model
different from those of \citet{Tsujimoto:2014} and \citet{Ishimaru:2015}.

This paper is structured as follows: we explain our methods in section~\ref{sec:2}, 
which contains one subsection summarizing the semi-analytic techniques that are used in our galaxy formation model
and another subsection explaining how the r-process (Eu) enrichment of the galaxy by NSM is modelled.
We show the results for a standard model in section~\ref{sec:3}, expand upon this result and discuss variations 
of the standard model in section~\ref{sec:4}. In section~\ref{sec:5}, 
we compare the effect of the assumption that the NS receive a kick at birth between BB galaxies and surviving satellites.
We end the paper with a summary and discussion section (section~\ref{sec:6}).

\section{Methods} \label{sec:2}

The build-up of the Galactic halo through merging building BBs is modelled using the Munich-Groningen
semi-analytic galaxy formation model \citep{Kauffmann:1999,Springel:2001,De-Lucia:2004,Croton:2006,De-Lucia:2007,
De-Lucia:2008,Li:2010,Starkenburg:2013}. As a backbone of this model, three of the six (A, B and C) high-resolution 
Milky Way mass Aquarius dark matter halo simulations are used \citep{Springel:2008}. Resolution level 2 is used 
($\sim 10^4 M_\odot$ per particle), because only Aquarius A is available at higher resolution. The semi-analytic model 
does not trace the evolution of individual elements explicitly, and assumes an instantaneous recycling approximation (IRA). 
Therefore, it does not accurately predict the abundance of elements that are produced on long timescales, such as Fe. 
Also, as we explain in section~\ref{sec:2.2}, the abundance of Eu as a function of cosmic time was put into the model 
explicitly for this study.

\subsection{The flow of baryons in the semi-analytic model}\label{sec:2.1}
\begin{figure}
\includegraphics[width=\columnwidth]{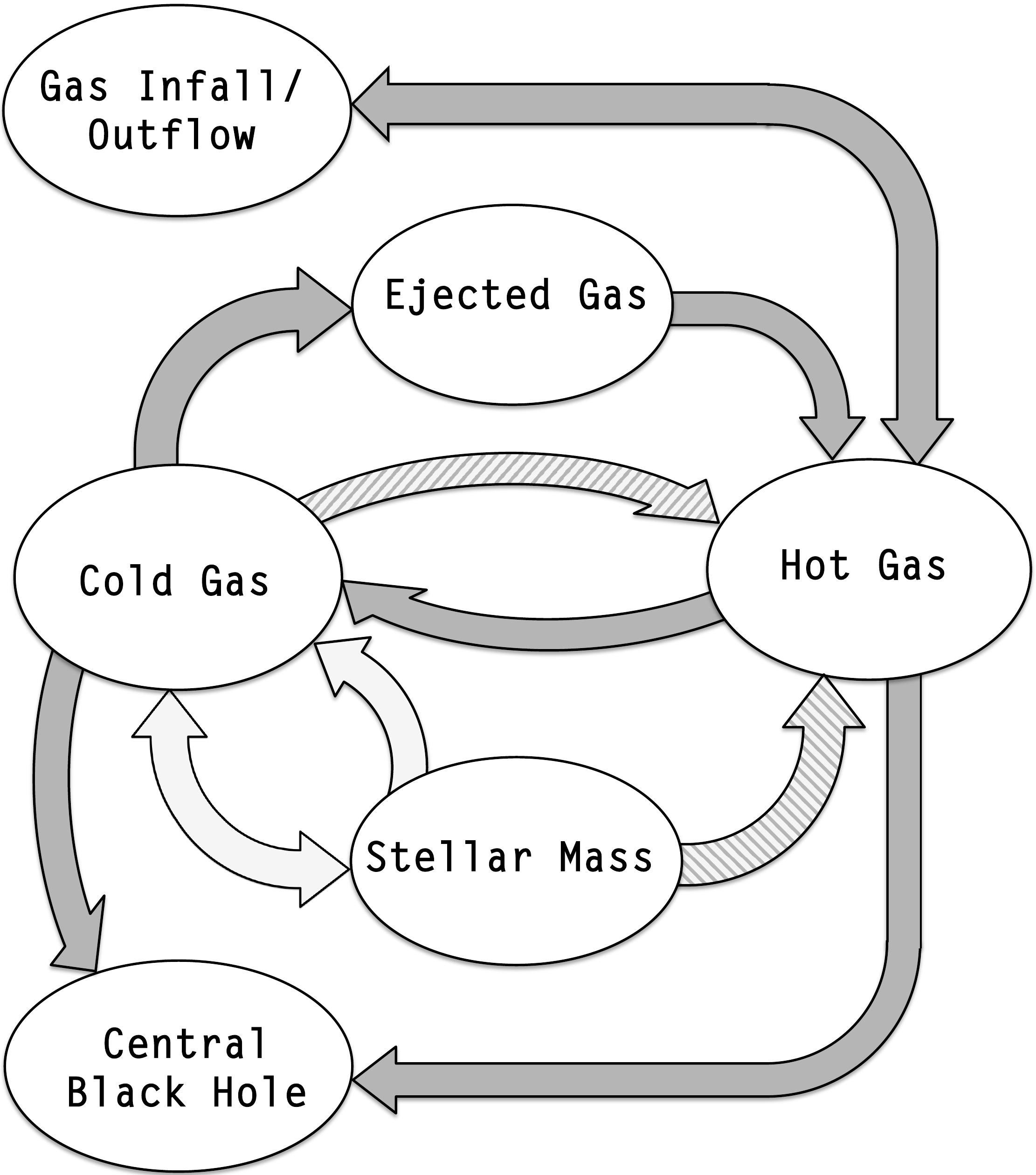}    
\caption{Flow of baryons in the semi-analytic model. The physical processes that 
drive the mass exchange are explained in the text. Dark filled arrows are only
allowed for main galaxies (type 0). Satellite galaxies only have Cold Gas
and Stellar Mass boxes. Two possible heating processes are behind the single hatched arrow from Cold Gas to Hot Gas,
although one of these adds baryons to the Hot Gas component of the main galaxy in the dark matter halo, 
a process that is only allowed for satellite galaxies. Since satellites do not have Hot Gas in our simulation, 
in case baryons are moved from the Stellar Mass or Cold Gas boxes in satellites to the Hot Gas box, it is 
the Hot Gas component of the main galaxy in the dark matter halo.
The dashed line from Stellar Mass to Hot Gas furthermore indicates that this process is not allowed for galaxies
with a viral mass above $5 \cdot 10^{10} M_\odot$.
Note that the arrow from Gas Infall/Outflow to Hot Gas points in both directions 
because a single prescription determines the direction of the arrow. The arrow from 
Cold Gas to Stellar Mass points in both directions to visualize the adopted instantaneous 
recycling approximation. }
\label{fig:2}
\end{figure}

As described in detail in the literature \citep[eg.][]{Starkenburg:2013}, the semi-analytic galaxy formation model that we 
use distinguishes between three different types of galaxies. Main galaxies in the centre of a main dark matter halo (type 0), 
satellite galaxies in the centre of a dark matter subhalo (type 1) and so called orphan galaxies that have lost their dark 
matter subhalo through tidal disruption (type 2). For type 0 galaxies, the flow of baryons in the model is visualized in 
Figure~\ref{fig:2}. In the next paragraphs, we will briefly summarize the physical processes that drive the mass exchange
between the different boxes baryonic mass can be in, since these also affect the route of Eu. As soon as a galaxy becomes
a satellite, the dark filled arrows in Figure~\ref{fig:2} are no longer allowed. One arrow with the dashed lines represents
two possible heating processes, of which one is only allowed for satellites.

In the top left corner of Figure~\ref{fig:2}, the infalling (pristine) gas or outflowing hot gas is shown. 
The amount of infalling/outflowing gas $M_\mathrm{i/o}$ is determined by the 
difference between the assumed (average) baryonic mass $\overline{M}_\mathrm{b}(M_\mathrm{vir},z)$ of the galaxy with virial mass $M_\mathrm{vir}$ at redshift $z$,  
taking reionization into account, and the baryonic mass in the galaxy, i.e. the sum of the mass in the other five elliptically shaped boxes of Figure~\ref{fig:2},
$M_\mathrm{b} = M_\mathrm{*} + M_\mathrm{cg} + M_\mathrm{hg} + M_\mathrm{eg} + M_\mathrm{bh}$.
\begin{equation}
M_\mathrm{i/o} = \overline{M}_\mathrm{b}(M_\mathrm{vir},z) - \sum_\mathrm{i} M_\mathrm{b,i}
\end{equation}
where the sum is taken over all (i) galaxies in the halo. $\overline{M}_\mathrm{b}(M_\mathrm{vir},z)$ is used as approximated by \citet{Gnedin:2000},
\begin{equation}
\overline{M}_\mathrm{b}(M_\mathrm{vir},z) = \frac{f_\mathrm{b} M_\mathrm{vir}}{\left[1+0.26 M_F(z)/M_\mathrm{vir}\right]^3}
\end{equation}
with the filtering mass $M_F(z)$ as described in Appendix~B of \citet{Kravtsov:2004}. A cosmic
baryon fraction $f_\mathrm{b}$ of 0.17 is assumed, consistent with the first-year WMAP results \citep{Spergel:2003},
and reionization is assumed to take place between redshift 15 and 11.5 \citep[see also][]{Li:2010}. 
Since only hydrogen is affected by reionization, Eu does not follow this arrow.
However, the process is shown for completeness.

Hot gas cools following the prescriptions described by \citet{De-Lucia:2004}, \citet{Croton:2006} and \citet{Li:2010}. The 
cooling rate is strongly dependent on the gas temperature and on its metallicity. The collisional ionization cooling curves of 
\cite{Sutherland:1993} are used to model these dependencies. Galaxies with a virial temperature $T_\mathrm{vir}(z)$
below the atomic hydrogen cooling limit ($T \sim 10^4$K) are prevented to cool gas in our model, since it is assumed 
to be very inefficient at early times.
We assume that the amount of Eu to the total metal budget of a galaxy that accounts for cooling is negligible at all times.
All metals (including Eu) that are contained in the fraction of the hot gas that is cooling are also cooled 
onto the cold disk of the galaxy.

From the cold gas box, baryons can flow to three different other boxes of Figure~\ref{fig:2} (or four if it is a main
galaxy). 
For main galaxies (type 0), the energy injected by SN, as described by \citet{Kauffmann:2000} is deposited into the ejected gas box of Figure~\ref{fig:2}.
\begin{equation}
\Delta M_\mathrm{reheated} = \frac{4}{3} \epsilon \frac{\eta_\mathrm{SN}E_\mathrm{SN}}{V_\mathrm{vir}^2} \Delta M_\mathrm{*} \label{SN_feedback}
\end{equation}
with $\Delta M_\mathrm{*}$ is the stellar mass formed, $V_\mathrm{vir}$ the circular velocity of the galaxy at the viral radius ($R_\mathrm{vir}$),
$E_\mathrm{SN} = 10^{51}$~erg the energy released per SN, and $\epsilon=0.05$ and $\eta_\mathrm{SN} = 8\cdot 10^{-3} M_\odot^{-1}$ 
are the feedback efficiency and the number of SNe respectively \citep{Li:2010}. 
We define the virial radius as the radius of a sphere with mass $M_\mathrm{vir}$ and mean interior density 
is 200 times the critical density for closure of the universe \citep{Navarro:1997aa}:
\begin{equation}
M_\mathrm{vir} = \frac{4}{3}\pi R_\mathrm{vir}^3 200 \frac{3 H^2}{8 \pi G} = \frac{100 H^2 R_\mathrm{vir}^3}{G}
\end{equation}
with H the Hubble parameter.
Smaller galaxies have shallower potential wells
and thus a lower $V_\mathrm{vir}$, which makes them more efficient reheaters. 

Two different physical processes are behind the arrow from cold gas pointing back to hot gas (highlighted with 
dashes), describing reheating of the gas. One is related to the radio activity of the active galactic nucleus (AGN) of the 
galaxy, as described by \citet{Croton:2006}, and should technically be considered a suppression of the cooling flow. 
This recipe is applied to all galaxy types. The other heating recipe again corresponds to the energy injected by SN,
equation~(\ref{SN_feedback}), but then only applied for satellite galaxies (types 1 and 2).
For those, the virial mass is taken to be the number of particles in the simulated galaxy times
the particle mass.
For both reheating recipes, all metals (including Eu) that are contained in the fraction of the cold gas 
that is reheated are moved to the hot gas phase of the host galaxy.

The baryons in the ejected gas box, $M_\mathrm{eg}$,
can be re-incorporated into the hot gas box, following the prescription of \citet{De-Lucia:2004}:
\begin{equation}
\dot{M}_\mathrm{re-inc.} = \gamma_\mathrm{eg} \cdot M_\mathrm{eg} \frac{V_\mathrm{vir}}{R_\mathrm{vir}}
\end{equation}
with $\dot{M}_\mathrm{re-inc.}$ the amount of gas that is reincorporated during one timestep,
and $\gamma_\mathrm{eg} = 0.5$ the ejected gas reincorporation efficiency \citep[][table~1]{Croton:2006}.
The ratio $R_\mathrm{vir}/V_\mathrm{vir}$ is proportional to the dynamical time of the galaxy \citep{Springel:2001}.
As for all the other processes, all metals (including Eu) contained in the fraction of the ejected gas 
that is re-incorporated are moved to the hot gas box.

Only main galaxies have ejected gas and hot gas components. Satellites transfer their ejected mass to their hosts (the mains).
This is indicated through the dashed line from Stellar Mass to Hot Gas in Figure~\ref{fig:2}.

Both cold gas and hot gas can be accreted onto the central black hole (BH).
With $M_\mathrm{bh}$, we denote the baryonic part of the BH mass. 
The change in this parameter during one timestep is indicated with an overdot.
From the hot gas box, 
quiescent accretion follows the empirical recipe described by \citet{Croton:2006}, i.e. the `radio mode':
\begin{equation}
\dot{M}_\mathrm{bh,r} = \kappa_\mathrm{AGN} \left(\frac{M_\mathrm{bh}}{10^{8} M_\odot}\right) \left(\frac{M_\mathrm{hg}}{0.1 \ M_\mathrm{vir}}\right) \left(\frac{V_\mathrm{vir}}{200 \ \mathrm{km \ s}^{-1}}\right)^3.
\end{equation}
Here, $M_\mathrm{hg}$ is the baryonic mass in the form of hot gas and $\kappa_\mathrm{AGN} = 7.5 \cdot 10^{-6} M_\odot \mathrm{yr}^{-1}$
the hot gas BH accretion rate \citep{De-Lucia:2007}. Cold gas accretion happens when a satellite galaxy merges with the central (main) galaxy. 
This recipe is also described by \citet{Croton:2006} and dubbed `quasar mode':
\begin{equation}
\Delta M_\mathrm{bh,q} = f_\mathrm{bh} \left(\frac{m_\mathrm{*} + m_\mathrm{cg}}{M_\mathrm{*} + M_\mathrm{cg}}\right) \frac{M_\mathrm{cg}}{1 + (280 \ \mathrm{km \ s}^{-1}/V_\mathrm{vir})^2 }
\end{equation}
with $M_\mathrm{*}$ and $M_\mathrm{cg}$ the stellar mass and mass in cold gas of the main galaxy and
$m_\mathrm{*}$ and $m_\mathrm{cg}$ the stellar mass and mass in cold gas of the satellite galaxy respectively.
$f_\mathrm{bh}=0.03$ is the cold gas BH accretion fraction. Again, all metals (including Eu) contained in the 
fraction of the hot or cold gas that is accreted are moved to the central BH box.

Last, but not least, we describe how stars are formed from cold gas and how metals are added to
the hot and cold gas components. We assume cold gas is transformed into stars in an exponential thin disc 
\citep[by the formalism of][]{Mo:1998}, which is assumed to extend to three scalelengths,
when its density is above a critical threshold \citep{Kennicutt:1989,Kauffmann:1996}:
\begin{equation}
\Sigma_\mathrm{crit} ( M_\odot \mathrm{pc}^{-2}) = 0.59 \  V_\mathrm{vir} (\mathrm{km \ s}^{-1}) / R_\mathrm{disc} (\mathrm{kpc}) 
\end{equation}
The star formation rate is then proportional to the amount of cold gas available. 
See also \citet{De-Lucia:2008} and \citet{Starkenburg:2013}.
Furthermore, during a galaxy merger, a collisional starburst recipe \citep{Somerville:2001} is triggered, where a fraction
\begin{equation}
e_\mathrm{burst} = \beta_\mathrm{burst} \left(\frac{m_\mathrm{*} + m_\mathrm{cg}}{M_\mathrm{*} + M_\mathrm{cg}}\right)^{\alpha_\mathrm{burst}}
\end{equation}
of the combined cold gas of the two galaxies ($m_\mathrm{cg} + M_\mathrm{cg}$) is turned into stars.
The parameters $\alpha_\mathrm{burst}$ and $\beta_\mathrm{burst}$ are set to be 0.7 and 0.56 respectively \citep{Croton:2006}.
Due to our adopted IRA and assumed \citet{Chabrier:2003} initial mass function (IMF), 43\% of the cold gas is returned immediately
to the cold gas component during star formation, hence the two-way pointing arrow in Figure~\ref{fig:2}.
Also, 43\% of the metals (including Eu) in the cold gas available for star formation are returned immediately 
to the cold gas, the other 57\% ends up in stars. 
After a starburst and instantaneous recycling, new metals (other than Eu, which is injected through a different channel - 
see section \ref{sec:2.2}) become available in the two gas phases. We follow the recipe of \citet{Li:2010}, who
suggested a two-state value of the fraction of metals deposited directly into the hot gas of the main galaxy in the
dark matter halo (which is the galaxy itself in case it is a type 0). For galaxies with a virial mass below 
$5 \cdot 10^{10} M_\odot$, 95\% of the metals are added to this box and 5\% to the cold gas box of the galaxy 
that produced the metals. Higher mass galaxies deposit all their new metals into the cold gas phase of their galaxy.
We assume a metal yield of 0.03 \citep{Croton:2006}.

\subsection{Adding the europium}\label{sec:2.2}
First we run the semi-analytic model once without any enrichment in europium, and record the total stellar mass
formed in each galaxy, as well as the maximum circular velocity of a particle in the potential well
of the galaxy at each time step. Then, assuming a NSM rate, we know in post-processing for each star formation episode 
how many NS mergers happen after a delay time picked out of an assumed delay time distribution (DTD). In our standard model,
we assume a NSM rate of $10^{-4}$ per $M_\odot$ stars formed. This assumption is based on the NS merger
rate inferred from gravitational wave (GW) observations \citep{Cote:2018aa}. 
It is about a factor $10$ higher than the rate assumed by \citet{Tsujimoto:2014}, who deduce a NSM rate of one per $\sim 1000$ core-collapse 
SNe\footnote{In our semi-analytic model we assume $8\cdot 10^{-3}$ core-collapse SNe per $M_\odot$ stars formed \citep{Li:2010}.}, 
and about a factor $100$ higher than that of \citet{Cescutti:2015}, 
who assume a NSM rate of a few times $10^{-6}$ per $M_\odot$ stars formed.

In our standard model, we assume a DTD proportional to $1/t$, with the first NSM occurring after 10~Myr, which we will 
abbreviate as 10 Myr~$+1/t$ in the remainder of this paper. As already mentioned in the introduction, 
this DTD is compatible with binary population synthesis studies \citep{Belczynski:2006}.

Assuming a NFW profile for the density of dark matter haloes at radius $r$ \citep{Navarro:1996}, 
\begin{equation}
\rho(r) = \frac{\rho_0}{\left(r/a\right)\left(1+r/a\right)^2}.
\end{equation}
where $\rho_0$ and $a$ are fit parameters, it can be shown that the escape velocity that a particle needs to have 
in order to escape from the centre of the potential well is
\begin{equation}
v_\mathrm{esc}(0) = \frac{\sqrt{2}}{0.465} V_\mathrm{max}, \label{eq:1}
\end{equation}
where $V_\mathrm{max}$ is the maximum of the circular velocity of the galaxy. At the scale radius $a$, 
the escape velocity is a factor $\sqrt{\ln(2)} = 0.833$ lower.

We assign a kick velocity to each NS binary at the formation time of the two NS, 
following the kick velocity distribution of \citet{Arzoumanian:2002}:
a two-component velocity distribution (Maxwellians) with characteristic velocities of 90 and 500 km/s,
with 40\% of the population being drawn from the distribution with the lowest characteristic velocity.
Since we assume that the system remains bound after receiving the kick, we assign
half the velocity picked from this distribution as the kick velocity of the binary.
We compare this velocity with the escape velocity from the centre of the galaxy 
hosting the NS binary (equation~\ref{eq:1}).
If the kick velocity is smaller than the escape velocity, we assume that 
the neutron star binary will be inside the galaxy at the time of the merger. 
Otherwise, the binary is discarded as a potential source of Eu enrichment.
We also investigate the effect of applying the low velocity Maxwellian distribution of 
\citet{Arzoumanian:2002} only, i.e. that with a characteristic velocity of 90 km/s. 
This assumption is hereafter referred to as the ``low kick'' model, which is
compared to the ``no kick'' (standard kick model) and ``with kick'' (assuming
the full kick velocity distribution) models.
The choice for the low kick model is motivated by the fact that many double neutron stars 
are found to have small NS kick velocities \citep{Tauris:2017aa}.
Furthermore, \citet{Verbunt:2017aa} have recently shown that 
a description with two Maxwellians, with distribution parameters 
$\sigma_1 = 77$ and $\sigma_2 = 320$ km/s, is significantly better than that of
\citet{Arzoumanian:2002}.

For those binaries that do not leave the host galaxy after receiving their kick,
we assume a europium yield of $1.5 \cdot 10^{-5} M_\odot$/NSM in our standard model.  
This is the upper limit of the Eu yield inferred from GW observations \citep{Cote:2018aa}
and slightly larger than what was assumed by \citet{Matteucci:2014,Matteucci:2015},
who assume Eu yields in the range $10^{-5} - 10^{-7} M_\odot$, but a factor $10$
smaller than the estimate of \cite{Komiya:2016aa}.
We assume that halo stars are only formed in BBs.
If a NSM happens when its hosting BB has already merged with the central galaxy, 
the gas enriched by this NSM might form new stars in the disk or inner spheroid,
but since this paper focusses on the halo component where no new star formation
takes place this is not relevant for our research.
Since the amount of Eu is negligible compared to the total NSM metal yield,
which is already added to its metal budget through the ``metals from stars to gas'' channel (see section~\ref{sec:2.1}),
we also do not add the Eu to the gas metal budget of the main galaxy's spheroid in this scenario.

Knowing exactly when a NSM happens in each galaxy,
we now re-run the semi-analytic model and incorporate the chemical evolution of Eu in the
ensemble of galaxies. The europium will be instantaneously mixed into the gas of each galaxy, 
which is used for the formation of new stars. Similar to the enrichment of the galaxy in other metals
(see section~\ref{sec:2.1}), we assume that 95\% of the newly produced Eu is mixed with the 
hot gas in the main galaxy in the dark matter halo if the galaxy that produced the Eu has
a virial mass below $5 \cdot 10^{10} M_\odot$, and 5\% is added to the cold gas budget of their 
own galaxy\footnote{In case the galaxy is not a satellite, the main galaxy in the dark matter halo is
also the galaxy itself.}. All Eu is added to the cold gas budget of the galaxy itself in case
it has a virial mass above $5 \cdot 10^{10} M_\odot$.

\subsection{Metallicities}

The total metallicity of a stellar population at each time step is traced by our model as the parameter $Z_\mathrm{stars}$, 
the ratio of mass in metals over the total mass in stars. Because of the IRA, this value can thus best be compared with 
the abundance of elements which are mainly originating in (short-lived) Supernovae type II, such as the $\alpha$-elements. 
In this paper we show predicted $\log[Z_\mathrm{stars}/Z_\odot]$ values of our model, with $Z_\odot = 0.0142$ the reference
metallicity \citep{Asplund:2009}.
These can thus be thought of as [Mg/H] values. Similarly, $\log [(X_\mathrm{Eu}/Z)_\mathrm{stars}/(X_\mathrm{Eu}/Z)_\odot]$
values, with $X_{\mathrm{Eu},\odot} = 3.6 \cdot 10^{-10}$ \citep{Asplund:2009}, can be thought of as [Eu/Mg].
That is why we chose to plot these abundance ratios against each other in Figure~\ref{fig:1}, instead of the more commonly 
used [Fe/H] and [Eu/Fe].

\section{Results for a standard model} \label{sec:3}

\begin{figure*}
 \includegraphics[width=\textwidth]{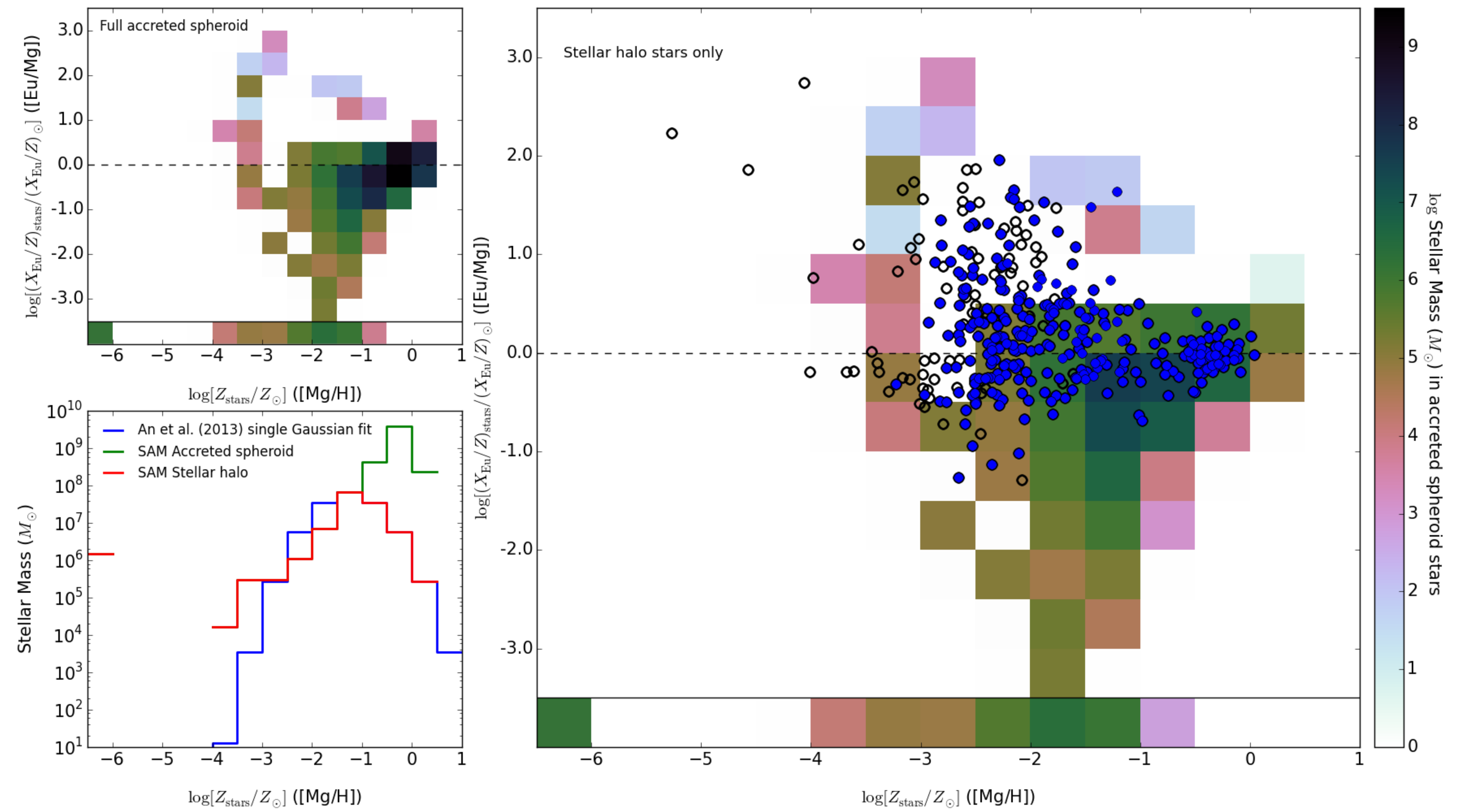}    
 \caption{Upper left panel: [Eu/Mg] vs. [Mg/H] map for the accreted spheroid of Aquarius halo A-2. Colors indicate the
  stellar mass in each bin on a logarithmic scale. The dashed line indicates the solar $X_\mathrm{Eu}/Z$
  ([Eu/Mg]) abundance, the solid line separates out the stars without any enrichment in Eu.
  The ones in the bottom leftmost bin also do not have other metals. These stars (still) exist in our model 
  because we neglect any kind of pre-enrichment from Population III stars.
  Lower left panel: MDF of this accreted stellar spheroid (green line), the single Gaussian fit
  to the observed IMF from \citet{An:2013} (blue line), normalized on the $-1.5\leq \log[Z_\mathrm{stars}/Z_\odot] < -1.0$
  bin from the model MDF, and the model MDF of stellar halo stars only (red line).
  Right panel: [Eu/Mg] vs. [Mg/H] map for the accreted stellar halo stars of Aquarius halo A-2 only,
  using the same colormap to indicate the stellar massas in each bin as in the left panel.
  Blue filled circles denote observed halo stars selected from the VizieR database \citep{Frebel:2010}
  with standard errors, blue open circles show halo stars selected from this database for which only 
  upper limits in [Eu/Mg] have been determined.  }
 \label{fig:3}
\end{figure*}

\subsection{The accreted spheroid}
\begin{figure*}
 \includegraphics[width=0.95\textwidth]{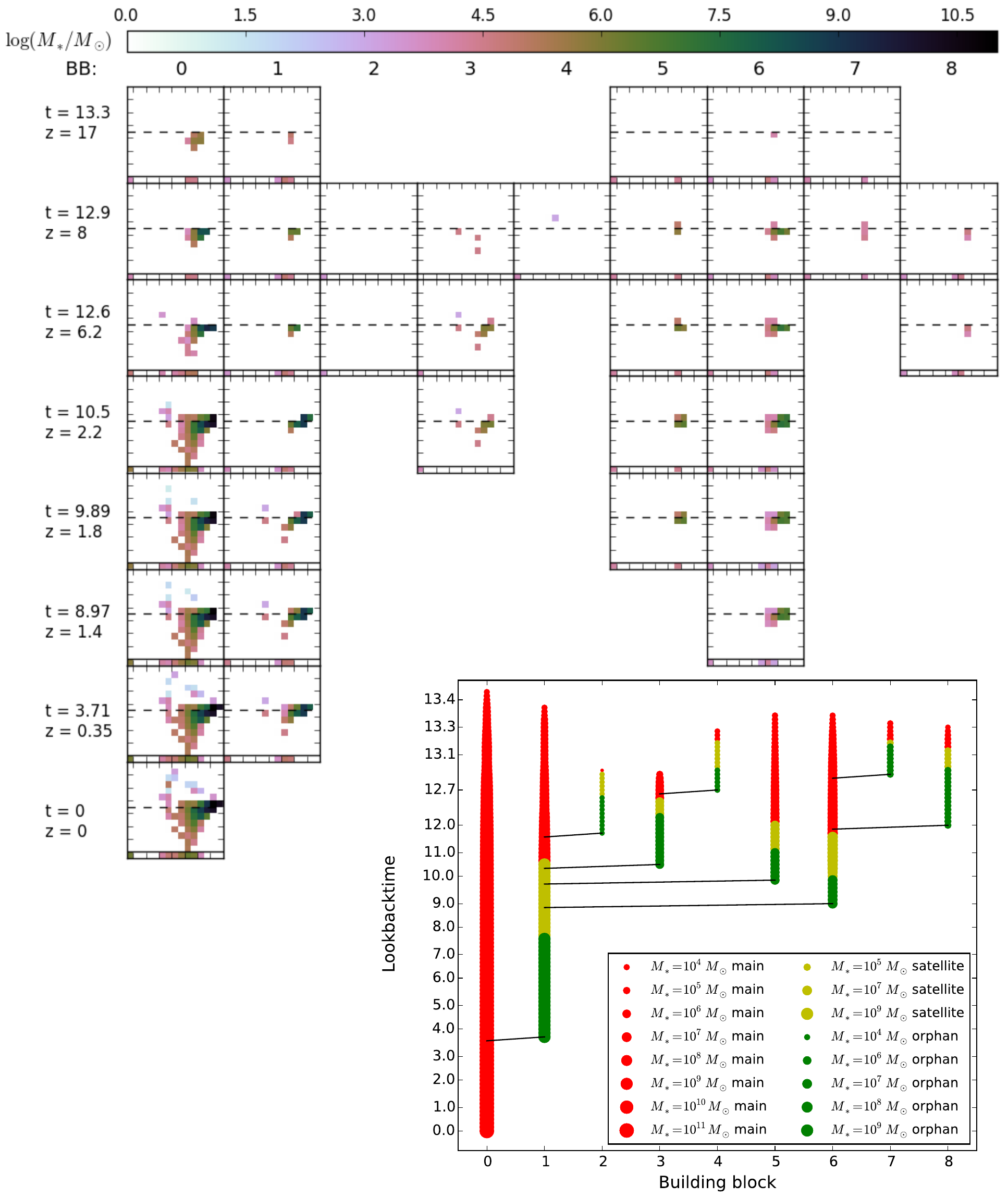}
 \caption{[Eu/Mg] vs. [Mg/H] maps for the simulated Milky Way galaxy in Aquarius halo A-2 (``building block'' 0)
 and one of its major building blocks (BB 1) and its building block galaxies (numbered 2 - 8).
 The galaxy merger tree is shown in the lower right-hand panel, 
 where the colors represent the different stages a building can be in, as explained in the beginning of 
 section~\ref{sec:2.1}: red for main galaxies, yellow for satellites and green for orphans. The size of the
 circles corresponds to the stellar mass as indicated in the legend. A black line indicates a galaxy merger.
 In the main figure, eight time snapshots are shown, with the corresponding lookback time ($t$) and redshift ($z$)
 shown on the left. The colormap indicates the logarithm of the stellar mass per bin (note that a
 different scale is used than in Figure~\ref{fig:3}). The
 maximum $\log_{10}$ of the stellar mass of the building blocks $0-8$ from left to right is 
 0:~11.3 (simulated Milky Way), 1:~9.3, 2:~3.5, 3:~6.5, 4:~3.9, 5:~6.7, 6:~7.9, 7:~4.9 and 8:~5.0, respectively.}
 \label{fig:4}
\end{figure*}

In the upper left-hand panel of Figure~\ref{fig:3}, we show the [Eu/Mg] versus [Mg/H] map 
predicted by our model for the accreted spheroid of Aquarius halo A-2, assuming that the NS do not receive
a kick velocity at their formation time (see section~\ref{sec:2.2}). Colors indicate the stellar mass
per bin, on a logarithmic scale, with darker colors representing more stars. 

The accreted spheroid contains bulge stars and halo stars. Their combined metallicity distribution function (MDF),
is shown with a green line in the lower left-hand panel of Figure~\ref{fig:3}. Since the stars selected from the VizieR database
and presented in Figure~\ref{fig:1} that we want to compare our model [Eu/Mg] vs. [Mg/H] maps with are all halo stars, 
we want to divide this accreted spheroid into a bulge part and a halo part. 
To do this, we use the same technique as \citet{van-Oirschot:2017aa}. The semi-analytic
model underestimates the number of accreted spheroid stars with metallicities $-3\leq \log[Z_\mathrm{stars}/Z_\odot] < -1$.
Therefore, we assume all stars in these and lower metallicity bins to be halo stars, and consequently
all accreted bulge stars to have metallicities $\log[Z_\mathrm{stars}/Z_\odot] > -1$. The fraction of halo stars in these
$\log[Z_\mathrm{stars}/Z_\odot] > -1$ bins of the accreted spheroid MDF is set to follow a single Gaussian fit to
the observationally determined MDF by \citet{An:2013}, which is normalized to the 
$-1.5\leq \log[Z_\mathrm{stars}/Z_\odot] < -1.0$ bin of the accreted spheroid MDF 
and shown with a blue line in the lower left-hand panel of Figure~\ref{fig:3}.
The red line is the resulting stellar halo MDF of the semi-analytic model (SAM).

The downscaling of the stellar mass in the $\log[Z_\mathrm{stars}/Z_\odot] > -1$ bins of the [Eu/Mg] vs. [Mg/H] map 
of accreted spheroid Aq-A-2 is done linearly for all age bins, and the resulting map is shown in the right-hand panel
of Figure~\ref{fig:3}. The same scale for the colormap is used as for the [Eu/Mg] vs. [Mg/H] map in the left panel.
The observed halo stars shown in Figure~\ref{fig:1} are overplotted with blue circles here (i.e. stars known to be
in satellite galaxies are left out). Open circles represent stars for which only upper limits are
available, filled circles represent stars which have standard errors in the determined [Eu/Mg] values.

It is a known shortcoming of the Munich-Groningen semi-analytic model that it underestimates the number of 
low metallicity halo stars. Therefore, it is not surprising that our model cannot explain the large number of 
observed stars with [Mg/H] < -1.5. More importantly, hardly any of the few low-metallicity stars that are predicted 
by our model, have [Eu/Mg] > 0 abundances, whereas these are observed in the Galactic halo.

In the next section we give an example of how the Galactic halo is built up from BBs, in order
to investigate if we can confirm the findings of \citet{Tsujimoto:2014}, \citet{Ishimaru:2015} and \citet{Komiya:2016aa} 
that Eu-enhanced halo stars naturally occur in a Galactic halo that is formed from merging subhaloes.
In the remainder of this paper we will not select out the accreted bulge stars
from modelled spheroids [Eu/Mg] vs. [Mg/H] maps, because we will focus our comparisons with data solely 
on the lower metallicities where we expect to have accreted halo stars only.
However, the same method could be applied as in Figure~\ref{fig:3} to end up with [Eu/Mg] vs. [Mg/H] maps
of accreted halo stars only.

\subsection{Building blocks of the Galactic spheroid}

In this paper, we consider the accreted component of the stellar spheroid, which is assumed to be built
from a few main progenitor galaxies, in agreement with the findings of many others \citep[eg.][]{Helmi:2002,
Helmi:2003,Font:2006,Cooper:2010,Gomez:2013}. In order to study the contribution of one of these large
BBs of the stellar spheroid to the [Eu/Mg] vs. [Mg/H] map in depth, we show a part of the galaxy merger tree 
of Aquarius halo A-2 in Figure~\ref{fig:4} (bottom right). Only the main (simulated Milky Way) galaxy
and one major BB are shown, respectively with numbers 0 and 1. BB 1 itself is built 
from four other BBs (labelled 2, 3, 5 and 6), of which some in turn are also built from building 
blocks. 

In Figure~\ref{fig:4} we also show the [Eu/Mg] vs. [Mg/H] maps of these BB galaxies
as a function of cosmic (lookback) time. Each panel contains a [Eu/Mg] vs. [Mg/H] map,
with successive time snapshots of the same BB vertically ordered. The labels of the axes 
are left out for clarity, but they are identical to those of Figure~\ref{fig:3}.
However, the chosen colormap which again indicates the stellar mass in each bin 
of a [Eu/Mg] vs. [Mg/H] map, is slightly different from the one used in Figure~\ref{fig:3}.
This is because Figure~\ref{fig:4} also contains the stars in the disc of the main (Milky Way) galaxy 
of Aquarius A-2 in its leftmost panels (hence this colormap extends to $10^{11} M_\odot$), whereas
Figure~\ref{fig:3} only contains accreted spheroid stars (and its colormap extends only to $10^{9.5} M_\odot$). 
The structure of the galaxy merger tree (bottom right) is matched in the main part of the figure. 
The [Eu/Mg] vs. [Mg/H] maps of BBs that have not yet formed any stars, or those of the BBs that 
merged with other BBs in a particular time step, are not shown. 

From this figure, it is clear that all except the lowest mass BB of this major BB (i.e. number
2, which has a maximum stellar mass of $10^{3.5} M_\odot$) form stars with Eu. This is a result of the assumed NSM rate
of $10^{-4}$ per $M_\odot$ stars formed, since that makes the threshold BB mass for forming stars which are
enriched in Eu $\sim 10^4 M_\odot$. Furthermore, we see from this figure that 
most of the few accreted spheroid stars with enhanced Eu abundances in this model (shown
above the dashed line in each panel), as well as those with low-metallicities 
($\log[Z_\mathrm{stars}/Z_\odot] < 1.5$) were not born in (BBs of) this major BB. 

Then of course the question arises, how can we explain the large number of halo stars
with [Eu/Mg] > 0 abundances? We will investigate the effect of different assumptions on the
NS kick, the NSM rate or the DTD in the next section, but it might very well be our assumption of 
instantaneous mixing that results in these low [Eu/Mg] abundances. Eu-enhanced stars naturally arise 
due to the inhomogeneity of gas enriched by NSM, as was also found on the subgrid level of the cosmological 
magnetohydrodynamical IllustrisTNG simulations recently \citep{Naiman:2018aa}.

\section{The effect of NS kicks, the NSM rate and the DTD} \label{sec:4}

\begin{figure}
 \includegraphics[width=\columnwidth]{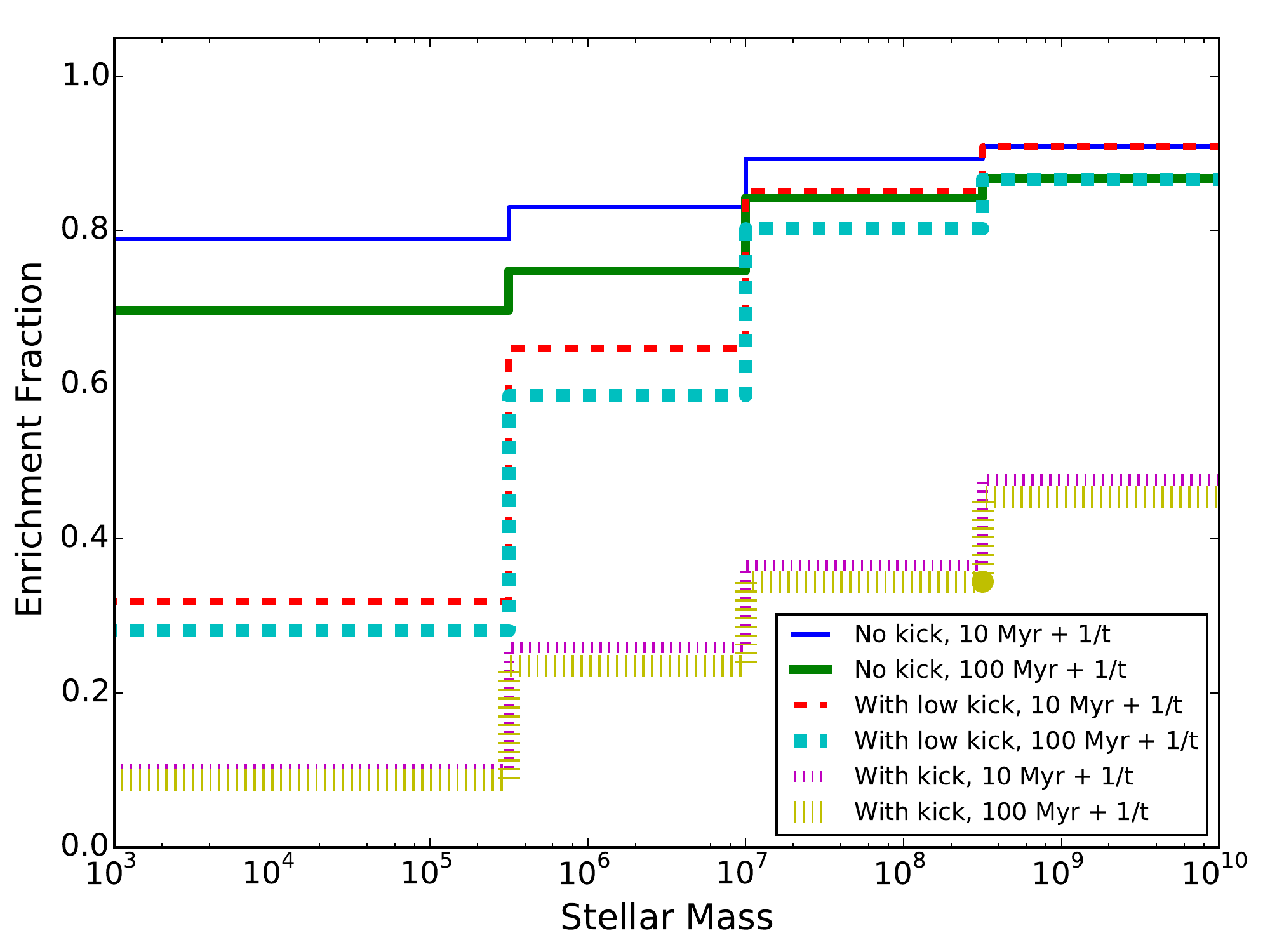}    
 \caption{Fraction of BBs per stellar mass bin that is enriched in Eu
 by NSM, for six different models of the binary NS population in the stellar halo. 
 The top (blue) solid line corresponds to our standard model (No kick, 10 Myr~$+1/t$)}. 
 \label{fig:5}
 \end{figure}
 \begin{figure*}
 \includegraphics[width=\textwidth]{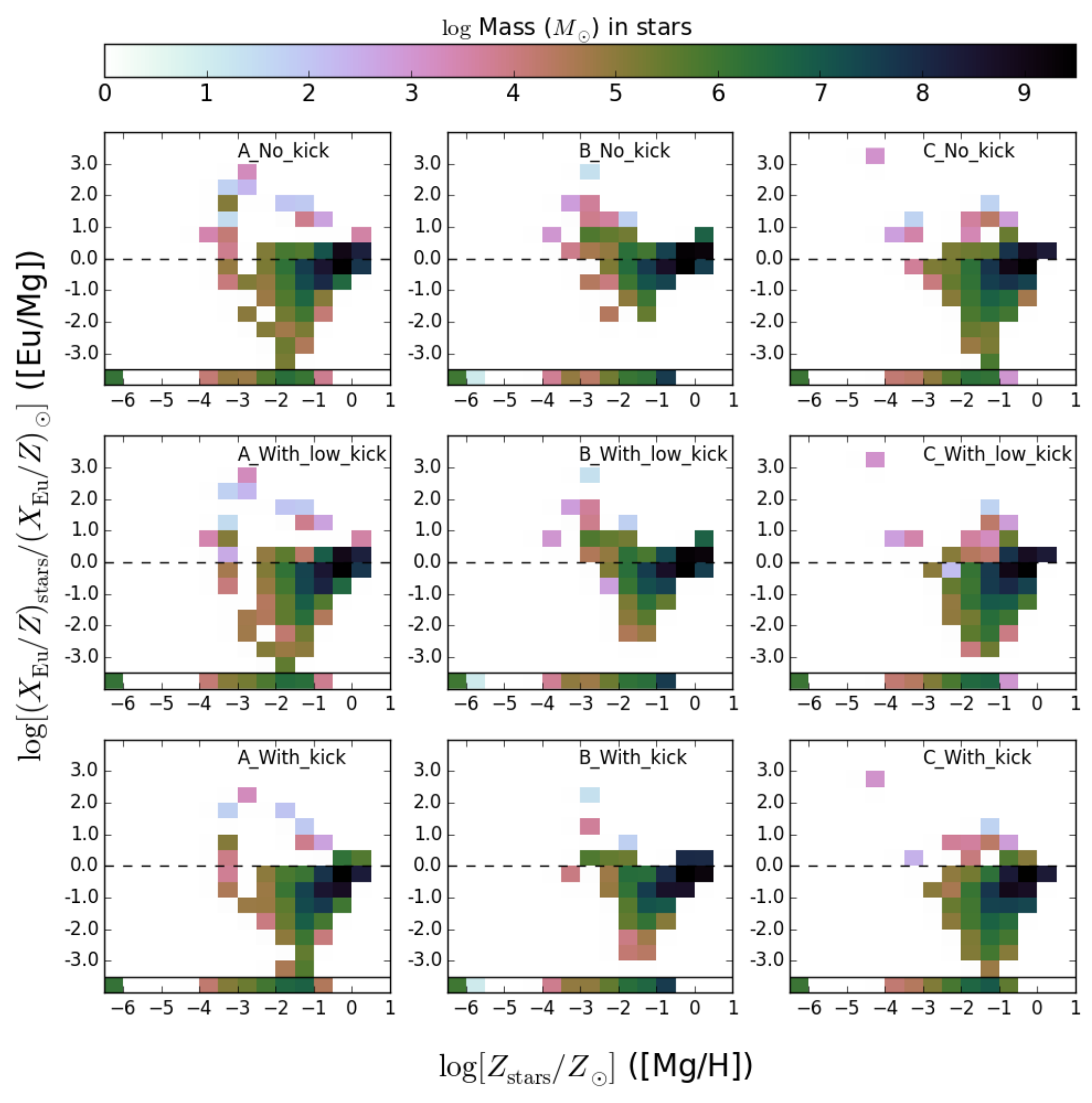}
 \caption{[Eu/Mg] vs. [Mg/H] maps for stellar spheroids in the Aquarius haloes A B and C (from left to right).
 The top, middle and bottom panels correspond to models in which the NS binary received no kick, a low kick or a higher
 kick, respectively.}
 \label{fig:6}
\end{figure*}

Although our standard model does predict some halo stars with high Eu-abundances for stars with low [Mg/H] values,
shown above the dashed line in the [Eu/Mg] vs. [Mg/H] maps presented in the previous section, compared
with observations (right-hand panel of Figure~\ref{fig:3}) the number density of Eu-enhanced stars 
predicted by the SAM applied to Aquarius halo A-2 is too low. Therefore, we will investigate variations of our
standard model in this section. 
Hereafter, we define enrichment in Eu to take place if all of the following three conditions are met:
\begin{itemize}
\item A NSM takes place in the BB.
\item The coalescence time of the NS binary is shorter than the timescale on which the BB merges with the central galaxy.
\item The NS kick velocity is smaller than the escape velocity of the BB.
\end{itemize}
The assumed NSM rate affects the first condition, the DTD affects the second condition
and the assumed binary NS kick velocity distribution affects the third condition. We will show the effect
of variations in all of these three variables.

\subsection{NS kick velocities} \label{sec:4.1}
In our standard model, we assume that NS binaries do not receive a kick, i.e. the third of the above 
conditions is always met. Therefore, introducing a kick velocity naturally decreases the fraction of 
Eu-enriched halo stars. To investigate by how much, we show in Figure~\ref{fig:5}
the fraction of BBs with Eu-enrichment by NSM as a function
of BB stellar mass, for three different assumptions on the NS kick velocity.
We also show the effect of two different assumptions on the DTD for each of these three models.

With our standard model assumptions (no kick) $\sim 90$\% of the highest mass BBs ($m_{*} > 10^{8.5} M_\odot$)
and $\sim 80$\% of the lowest mass BBs ($m_{*} < 10^{5.5} M_\odot$) is enriched in Eu.
Assuming longer delay times through a 100 Myr~$+1/t$ DTD 
in this case give $5-10$\% lower enrichment fractions, because fewer NS binaries merge
before the BB merges with the central galaxy in that model.
Introducing the 2-peak \citet{Arzoumanian:2002} kick velocity distribution,
drastically decreases this fraction, to less than 50\% of the highest mass BBs  
and to only 10\% of the lowest mass BBs.
This result is almost identical if we take a 100 Myr~$+1/t$ DTD instead of our standard 10 Myr~$+1/t$.
The enrichment fractions corresponding to the low kick models (only the low velocity peak of the \citet{Arzoumanian:2002} distribution)
fall in between the other lines for low mass BBs, and overlap with the lines for no NS kick for the highest mass BBs 
because these highest mass BB have escape velocities which are always larger than these low NS kick
velocities.

\begin{figure*} 
\includegraphics[width=\textwidth]{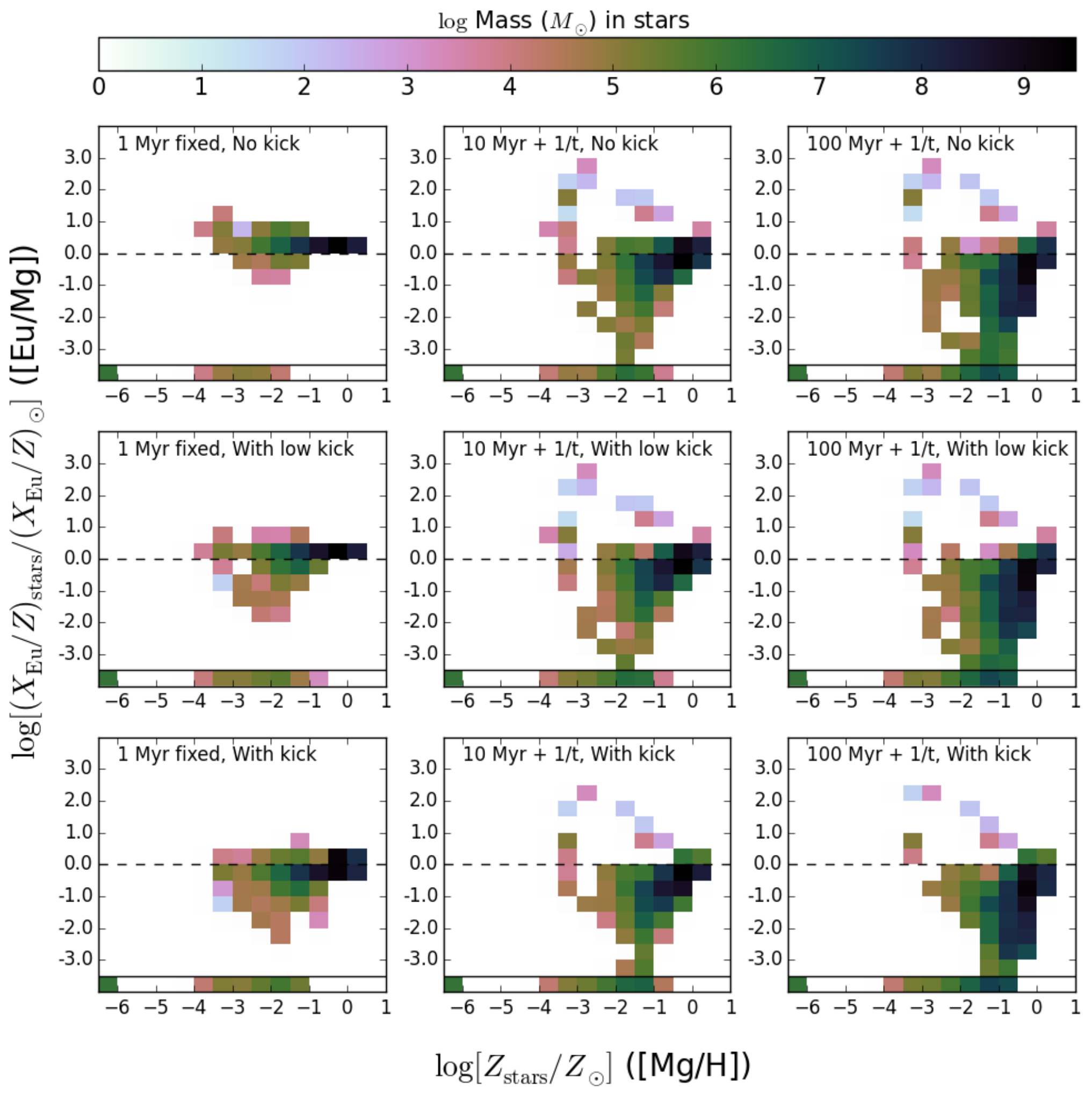} 
\caption{[Eu/Mg] vs. [Mg/H] maps of stellar spheroid Aq-A-2, for three different assumptions on the DTD:
1 Myr fixed, 10 Myr $+1/t$ and 100 Myr $+1/t$ from left to right. The tree different assumptions on the
NS kicks are again ``no kick'', ``low kick'' and ``with kick'' from top to bottom. }
\label{fig:7}
\end{figure*} 

\begin{figure*}
 \includegraphics[width=\textwidth]{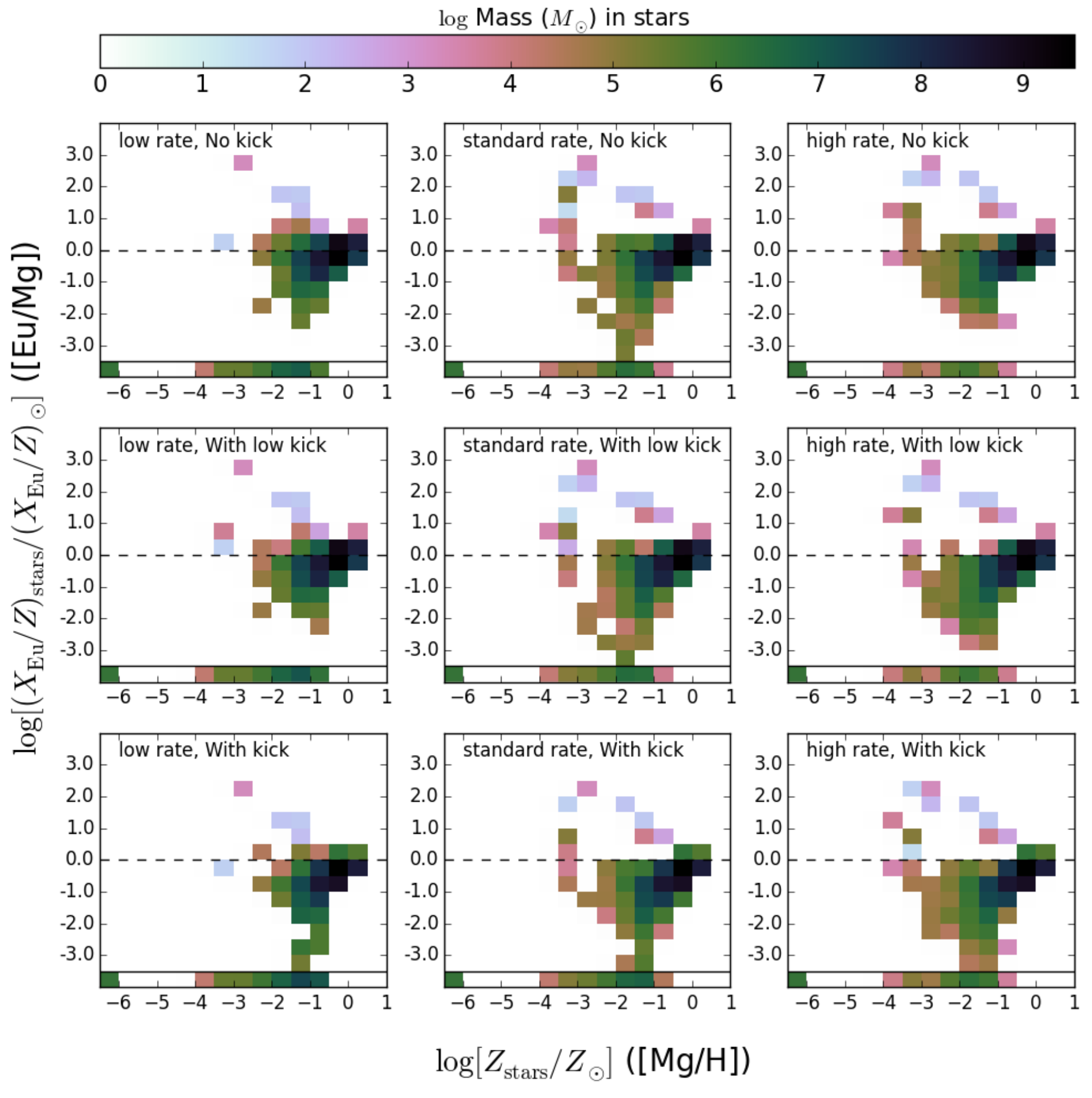}
 \caption{[Eu/Mg] vs. [Mg/H] maps of stellar spheroid Aq-A-2, for three different assumptions on the NSM rate: 
 low ($10^{-5} M_\odot^{-1}$), standard ($10^{-4} M_\odot^{-1}$) and high ($10^{-3} M_\odot^{-1}$) from left to right,
 and corresponding assumed Eu yields of $1.5 \cdot 10^{-4} M_\odot$/NSM, $1.5 \cdot 10^{-5} M_\odot$/NSM and $1.5 \cdot 10^{-6} M_\odot$/NSM. 
 The tree different assumptions on the NS kicks are again ``no kick'', ``low kick'' and ``with kick'' from top to bottom. 
 In all these models, the assumed DTD is 10 Myr~$+1/t$.}
 \label{fig:8}
\end{figure*} 

In Figure~\ref{fig:6}, the [Eu/Mg] vs. [Mg/H] maps are shown for the full accreted spheroids of Aq-A-2, 
Aq-B-2 and Aq-C-2 from left to right respectively. The top, middle and bottom panels show the difference 
between assuming that the NS binary receives no kick, a low kick, or a higher kick. A DTD of 10 Myr~$+1/t$ is assumed. 
Figure~\ref{fig:6} shows that there is indeed a larger Eu-enrichment in case the NS do not receive a kick.
Roughly, the distribution of stars in these [Eu/Mg] vs. [Mg/H] maps shifts 0.5 dex downwards when we assign  
a NS kick velocity. Furthermore, this figure gives an impression of the stochastic scatter between the Aquarius haloes. 
One cannot expect to find a galaxy in one of the Aquarius haloes that is an exact replica of our Milky Way, because 
only six simulations were done (of which we study three in this paper). 
In particular, our semi-analytic model ``no kick'' and ``with low kick'' applied to Aquarius halo B results in 
an [Eu/Mg] vs. [Mg/H] map that is in very good agreement with the observations. The majority of r-process enriched stars
in this halo originate in a single relatively massive ($10^{7.45} M_\odot$) BB, in which a relatively large starburst
occurred at late times ($10$\% of its stars were formed in a burst about $9.5$ Gyr ago, 
whereas most halo stars were formed about $11$ Gyr ago).
Our model thus predicts quite some variation between galaxies in terms of r-process enrichment.

\subsection{Delay time distribution} \label{sec:4.2}
The effect of assuming a different DTD on the [Eu/Mg] vs. [Mg/H] map of stellar spheroid Aq-A-2 
is shown in Figure~\ref{fig:7}. 
A constant delay time of 1 Myr may be unrealistically short to be
compatible with population synthesis studies \citep{Belczynski:2006}, but this is what has been assumed
before to explain Eu abundances in the Milky Way galaxy with a NSM enrichment scenario \citep{Matteucci:2014}.
The models in which a fixed delay time of 1 Myr is assumed yield
[Eu/Mg] vs. [Mg/H] maps where most stars have [Eu/Mg]$\sim 0$ with little scatter, since NSM are
treated almost like SNe Type II with this assumed DTD and fairly high NSM rate.
The bulk of the stars is shifted to higher [Eu/Mg] abundances by about $\sim$0.5 dex compared to our standard model (10 Myr~$+1/t$).
Thus the effect is similar to that of assuming a kick velocity for the binary NS.

Assuming a longer DTD on the other hand (100 Myr~$+1/t$), results in the majority of stars being spread
out over the low [Eu/Mg], high  [Mg/H] abundance bins (eg. $0\gtrsim$[Eu/Mg]$\gtrsim -2$ for $0\gtrsim$[Mg/H]$\gtrsim -1$),
which is clearly not where the majority of the observed halo stars are (see Figures~\ref{fig:1} and \ref{fig:3}).

\subsection{NSM rate} \label{sec:4.3}
Three different NSM rates are assumed in Figure~\ref{fig:8}. These are combined with Eu yield assumptions
such that the Galactic disc stars have solar Eu abundances at [Mg/H]=0 in our models,
i.e. the production of the NSM rate and the Eu yield always equals $1.5 \cdot 10^{-9}$/NSM. With high NSM rates
($10^{-3} M_\odot^{-1}$) and low Eu yields ($1.5 \cdot 10^{-6} M_\odot$/NSM), the effect of continuous Eu production is studied, similar to how it would be in 
a supernova enrichment scenario. We also investigate a scenario in which NSM are a factor 10 
rarer than in our standard model ($10^{-5} M_\odot^{-1}$), but yield a large amount of Eu ($1.5 \cdot 10^{-4} M_\odot$/NSM).

In the left-hand panels of Figure~\ref{fig:8}, we show the models with an increased Eu yield and correspondingly lower NSM rate, 
compared to our standard model. Here, only the most massive BBs are
expected to form Eu-enriched stars. This can also be seen from the number of stars
without any Eu, that end up in the lowest [Eu/Mg] bin ([Eu/Mg]$<-3.5$, below the solid line) of each [Eu/Mg] vs. [Mg/H] map.
There are much more of these in the left-hand panels of Figure~\ref{fig:8} 
(low NSM rate) than in the right-hand panels (high NSM rate).
The difference between the middle and the right-hand panels seems not very large.

\begin{figure*}
 \includegraphics[width=0.95\textwidth]{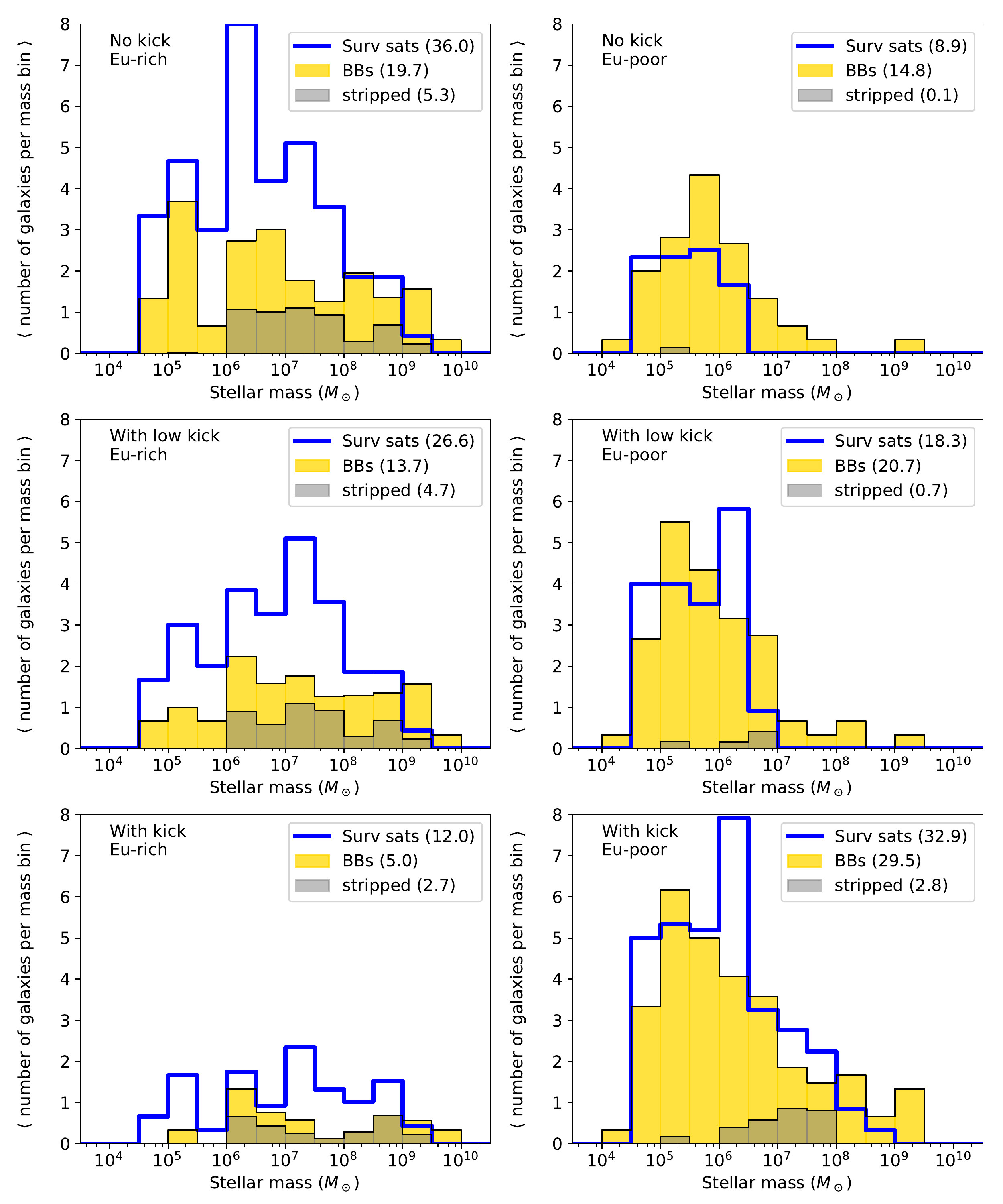}
 \caption{Average number of building blocks (filled gold histograms) and surviving satellites (transparent histograms with thick blue edges) 
 with more than 4 star formation snapshots (see text) per mass bin in two different categories: Eu-rich (left-hand panels) 
 and Eu-poor (right-hand panels), for three different assumptions on the NS kick: no kicks (top panels), low velocity kicks (middle panels), 
 or higher kick velocities (bottom panels). See section~\ref{sec:5} for a definition of how the Eu-rich and Eu-poor categories are defined.  
 The average is taken over the three Aquarius haloes A, B and C. Material stripped from a surviving satellite
 is partially counted as building block and the fraction of the initial satellite mass that is not stripped is contributes as
 surviving satellite. The dark shaded filled histogram shows how this material contributes to the building block distribution. 
 The total average number of building blocks and surviving satellites per category is shown in brackets in the legend.}
 \label{fig:9}
\end{figure*}

\section{Comparing BBs with surviving satellites} \label{sec:5}

Defining galaxies that contain stars with [Eu/Mg]>0 to be Eu-rich, and those with some Eu-enriched stars 
but no stars above the solar [Eu/Mg] abundance as Eu-poor,
we show the average number of galaxies in these two categories as a function of galaxy
stellar mass in Figure~\ref{fig:9}, for the three different assumptions on the NS kick. 
The average is taken over the three Aquarius haloes A, B and C.
We distinguish BBs (filled gold histograms) and surviving satellites (transparent histograms with thick blue edges) 
to investigate the effect of the timescale on which the galaxies merge with the central galaxy:
surviving satellites per definition have not (yet) merged with the central galaxy in the simulation.
If a galaxy is not fully disrupted, it is counted as partially stripped and divided beween the BB
and surviving satellite categories according to the fraction of the total stellar mass that is stripped. 
For example, if 70\% of its initial mass is stripped, the galaxy is counted as 0.7
BB stripped material and 0.3 surviving satellite galaxy. 

We do not trust our model's prediction for galaxies with a stellar mass below a few times $10^4 M_\odot$
due to the limiting resolution of our simulation. Largely overlapping with these, the galaxies with less 
than five star formation snapshots are left out of Figure~\ref{fig:9}, because the neglect of pre-enrichment 
from the very first generation of stars in our model is most clear in these galaxies \citep{Starkenburg:2013}.
This can be thought of as a mass cut since more massive galaxies sustain star formation for more extended periods
and show star formation in more than five snaphots. The number five is a consequence of the time resolution of 
our code, which is redshift dependent but in the order of tens to hundreds of Myr. 
The ultra-faint dwarf galaxy Reticulum II (Ret II), which was recently discovered to be extremely enriched 
in r-process elements \citep{Ji:2016,Ji:2016a}, falls below this mass resolution limit \citep{Simon:2015}.
Although we assume a NSM rate of only one per $10^4 M_\odot$ stars formed, 
due to Poisson scatter galaxies with stellar masses as low as that of Ret II can have Eu-rich stars 
in the NSM r-process enrichment scenario. In the left-hand panels of Figure~\ref{fig:9}, it can be seen that
this is most likely to be the case if the NS binary receives no kick, or a kick with a low velocity.

The fact that there is a large difference between the number of stars in the three categories between the
three assumptions on the NS kick, is related to our definition of Eu-rich stars, i.e. those with [Eu/Mg]>0.
By introducing a kick velocity, the distribution shifts to lower [Eu/Mg] abundances,
by about $\sim$0.5 dex.
Those galaxies with [Eu/Mg] abundances just above zero move to values just below zero when a kick is introduced,
which makes them fall into a different category.

Figure~\ref{fig:9} also predicts a difference between the Eu enrichment of the
stellar halo of the Milky Way and that of its surviving satellite galaxies.
Surviving satellites more often are expected to have Eu-rich stars than those that are fully disrupted.
An explanation for this can be found in both the time the galaxy takes to merge with the central galaxy, and in
the average timespan of star formation, that is longer for 
surviving satellite galaxies than for disrupted BBs \citep{van-Oirschot:2017}.
Galaxies with a longer star formation timespan for a similar final stellar mass have a lower star formation
rate, a slower build-up of the metallicity (Mg) budget due to Type-II supernovae, and correspondingly 
more stars with high [Eu/Mg] abundance. A similar argumentation
is used to explain the high alpha-over-iron element abundances in the stellar halo compared to that of surviving
satellite galaxies \citep[eg.][]{Font:2006,Font:2006a,Geisler:2007,Robertson:2005}.

In the top left panel of Figure~\ref{fig:9} we see that in our standard model, 27\% of the modelled 
Milky Way spheroid stars with [Eu/Mg]>0 comes from BBs that still have a surviving counterpart.
For the models ``with kick'' this ratio is as high as 54\%, as can be seen in the bottom left panel of Figure~\ref{fig:9}.
However, it is unlikely that many stars with high Eu abundances in the Milky Way halo are stripped from surviving
satellites, since the Eu-enhanced are local field halo stars for which no known stream or parent is known.
Furthermore, these stars are [$\alpha$/Fe] enhanced, which the satellite galaxies that have survived are not.

\section{Summary and discussion} \label{sec:6}

In this paper, we have modelled the europium production by neutron star mergers in the halo of the Milky Way
with the Munich-Groningen semi-analytic galaxy formation model. In particular, we have investigated the 
effect of the kick velocity the NS binary receives upon its formation, which may lead to a NS merger
outside the BB galaxy hosting the double NS. Although this may lead to Eu enrichment in the inner spheroid
or disk of the main galaxy in the simulation, in our models this situation will not lead to enrichment
of the stellar halo, which is assumed to be formed from merging BBs only. The effect is dependent on the
NS kick velocity and the escape velocity of the hosting BB. The NS kick velocity influences the amount of enrichment 
of the Galactic spheroid in Eu: in [Eu/Mg] vs. [Mg/H] stellar density maps, stars have $\sim 0.5$ dex lower [Eu/Mg]
abundances if a NS kick drawn from the \citet{Arzoumanian:2002} velocity distribution is assumed (Figure~\ref{fig:6}).
Of the BBs with a stellar mass below $10^{5.5} M_\odot$, only 10\% will be enriched in Eu when a NS kick is assumed,
compared to $\sim 80$\% when no NS kick is assumed (Figure~\ref{fig:5}). For more massive BBs the enrichment is
always larger, as these have deeper potential wells from which NS binaries have to escape. 

We assumed the velocity of the center of mass of the binary to be half the kick velocity, which
is slightly oversimplifying the effect of the binary evolution. Neglecting the effect of the kick after the
supernova of the first star, and assuming the star that undergoes the second supernova drags along the
other star, which is assumed to be of equal mass, we arrive at the factor 2. Once a complete binary stellar 
evolution model is combined with the semi-analytical galaxy formation model, a more accurate value for the 
system velocity after the supernova of the second star can be assumed.

Our simulations indicate that our standard assumption on the DTD (proportional to $1/t$, with the first merger 
occurring after 10 Myr) results in [Eu/Mg] vs. [Mg/H] maps which match that of the observed stellar halo population reasonably well, 
for stars with metallicities [Mg/H]$\geq -1.5$. 
Although we confirm the result of \citet{Tsujimoto:2014}, \citet{Ishimaru:2015} and \citet{Komiya:2016aa} 
that a stellar
halo built from merging building block galaxies naturally hosts stars with high Eu abundances at low metallicities,
our model predicts too few low-metallicity spheroid stars.
The MDF bins with metallicities between $-3\leq \log[Z_\mathrm{stars}/Z_\odot] < -1$ contain too few
stars compared to the single-Gaussian fit to the observed MDF of the Galactic halo by \citet{An:2013}.
It is exactly in this metallicity range, where surprisingly many stars are found to be enriched in 
Eu \citep{Frebel:2010}. Contrary to the findings of others, eg. \citet{Komiya:2016aa} or \citet{Naiman:2018aa}, 
we do not find many Eu-rich stars (i.e. those with [Eu/Mg]>0) at low metallicity in our model of the Galactic stellar halo, 
although for Aquarius halo B there are stars with [Eu/Mg]>0.5 at 
$-3\leq \log[Z_\mathrm{stars}/Z_\odot] < -1.5$ in our simulations when we assume that NS binaries do not receive a kick velocity
or receive a low kick velocity. 

A delay time proportional to $1/t$, but with the first merger occurring after 100 Myr (instead of 10 Myr as assumed in our standard model)
results in too many stars with high metallicity and low Eu abundance (Figure~\ref{fig:7}). A shorter, fixed delay time of 1 Myr, 
which \citet{Matteucci:2014} concluded to be necessary for a pure NSM-enrichment scenario for the Galactic stellar halo,
results in [Eu/Mg] vs. [Mg/H] maps with little scatter around [Eu/Mg]$\sim 0$.
Furthermore, population synthesis studies favor longer delay times \citep{Belczynski:2006}.
Reducing or neglecting the NS kick velocity has a similar effect as assuming a shorter delay time, with respect to
increasing the Eu-abundances of the modelled halo stars.
To enhance the r-process enrichment of a simulated galaxy, one could either increase the Eu yield per NSM 
or assume a higher NSM rate. 
In Figure~\ref{fig:8} we resolve this degeneracy in high NSM rate and low r-process yield versus low rate and high yield. 
We find that assuming a lower NSM rate and higher Eu yield compared to our standard model results in a [Eu/Mg] vs. [Mg/H] map 
that is less in agreement with observations of Eu-enhanced halo stars.
A higher NSM rate and lower Eu yield on the other hand, does not solve the issue 
that there are too few Eu-enhanced halo stars with $\log[Z_\mathrm{stars}/Z_\odot] ([Mg/H]) < -1.5$ in our standard model.

Although we do not assume instantaneous recycling of Eu, our neglect of the short ($\sim 10$~Myr) but nonzero
recycling time of alpha-elements such as Mg could partly explain the discrepancy with the observed stellar halo stars.
It is more likely however, that the absence of these halo stars with high Eu abundances at low metallicities 
indicates that our assumption of instantaneous (i.e. homogeneous) mixing is incorrect.
From the chemical signatures of the metal-poor stars in dwarf galaxies there have been hints of inhomogeneous mixing processes,
for instance of SNe Type Ia products \citep[see, for example][]{Venn:2012aa,Starkenburg:2013aa,Jablonka:2015aa}, although
the energies and outflow of new material of SN Ia are quite different from the energies and outflow of NSM products.
This paper provides another line of evidence that the gas in dwarf galaxies and BBs is probably inhomogeneously mixed at early times.
Furthermore, there are several works that suggest that the IMF in ultra-faint dwarf galaxies 
can be very different from the standard one \citep[e.g.][]{Geha:2013aa}. Since a very different IMF can change the 
neutron star density in such a stellar population, this leads to even more inhomogeneity in r-process enrichment.

An alternative conclusion one could draw, is that an additional production mechanism for r-process elements is necessary to explain
the discrepancy between our models and the observations. This mechanism could enhance halo stars in r-process elements
already at early times, when the metallicity of the halo stars is low and many building blocks of the stellar halo still have shallow potential 
wells from which NS binaries easily escape after receiving a standard kick velocity. 
This scenario is supported by the recent observation of a only mildly r-process enhanced star in the ultra-faint 
dwarf galaxy Tucana III \citep{Hansen:2017aa}, contrary to the highly r-process enhanced stars observed in Ret II.

Finally, the star formation history of (the BBs of) the stellar halo could play a role, 
since we found that Aquarius halo B is more in line with observations than haloes A and C (see Figure~\ref{fig:6}, which 
gives an impression of the stochastic scatter between the Aquarius haloes). 
In a comparison between the Eu-enrichment in our modelled BB galaxies of the Galactic spheroid and the surviving 
satellite population, we also find that the surviving satellites on average more often have stars with [Eu/Mg]>0 abundances 
than the fully disrupted BBs and that low metallicity spheroid stars 
with high Eu abundances are often stripped from a satellite galaxy with a still surviving counterpart (Figure~\ref{fig:9}). 
However, this is unlikely to be the case in the Milky Way halo, since there is no observational evidence supporting it. 

\citet{van-de-Voort:2015} found that stars with low [Fe/H] and high [Eu/Fe] are at large galactocentric radius
in a cosmological zoom-in simulations of a Milky Way-mass galaxy, but they are formed in-situ at high redshift and moved
there through dynamical effects. Furthermore, \citet{Naiman:2018aa} find no correlation between europium abundances 
and assembly history, in Milky Way-like galaxies in the IllustrisTNG simulations.

\section*{Acknowledgements}
The authors are indebted to the Virgo Consortium, which was responsible for designing and running the halo simulations of the Aquarius Project 
and the L-Galaxies team for the development and maintenance of the semi-analytical code. In particular, we are grateful to Gabriella De Lucia and 
Yang-Shyang Li for the numerous contributions in the development of the code.
We also thank Amina Helmi for valuable comments.

\bibliographystyle{mnras}
\bibliography{bibliography} 
\label{lastpage}
\end{document}